\documentclass[twocolumn,trackchanges]{aastex63}
\usepackage{booktabs}
\usepackage{gensymb}
\newcommand{\kstest}{$-2.9^{+1.2}_{-1.0}$} %bootstrap errors
\newcommand{\bayesage}{$54.3^{+3.8}_{-2.0}$\ }
\newcommand{\bayesmass}{$7.1^{+0.1}_{-0.2}$\ }
\newcommand{\M}{$M_{\odot}\ $}

\turnoffeditone

\begin{document}

\title{The Masses of Supernova Remnant Progenitors i n M33}

\shorttitle{M33 SNR Progenitor Masses}
\shortauthors{Koplitz et al.}

\correspondingauthor{Brad Koplitz} 
\email{bmk12@uw.edu}

\author[0000-0001-5530-2872]{Brad Koplitz}
\affil{Department of Astronomy, Box 351580, 
University of Washington, Seattle, WA 98195, USA}

\author[0000-0002-2630-9490]{Jared Johnson}
\affil{Department of Astronomy, Box 351580, 
University of Washington, Seattle, WA 98195, USA}

\author[0000-0002-7502-0597]{Benjamin F. Williams}
\affil{Department of Astronomy, Box 351580, 
University of Washington, Seattle, WA 98195, USA}

\author[0000-0002-4652-5983]{Mariangelly D{\'i}az-Rodr{\'i}guez}
\affil{Department of Physics, Florida State University,
77 Chieftan Way, Tallahassee, FL 32306, USA}

\author{Jeremiah W. Murphy}
\affil{Department of Physics, Florida State University,
77 Chieftan Way, Tallahassee, FL 32306, USA}

\author[0000-0003-3252-352X]{Margaret Lazzarini}
\affil{California Institute of Technology, 
1200 E California Blvd., Pasadena, CA 91125, USA}

\author[0000-0001-8878-4994]{Joseph Guzman}
\affil{Department of Physics, Florida State University,
77 Chieftan Way, Tallahassee, FL 32306, USA}

\author[0000-0002-1264-2006]{Julianne J. Dalcanton}
\affiliation{Center for Computational Astrophysics, Flatiron Institute, 
162 Fifth Ave, New York, NY 10010, USA}
\affil{Department of Astronomy, Box 351580, 
University of Washington, Seattle, WA 98195, USA}

\author[0000-0001-8416-4093]{Andrew Dolphin}
\affil{Raytheon Technologies,
1151 E. Hermans Road, Tucson, AZ 85706, USA}
\affil{Steward Observatory, University of Arizona,
933 N. Cherry Avenue, Tucson, AZ 85719}

\author[0000-0001-7531-9815]{Meredith Durbin}
\affil{Department of Astronomy, Box 351580, 
University of Washington, Seattle, WA 98195, USA}

\begin{abstract}

Using resolved optical stellar photometry from the Panchromatic Hubble Andromeda Treasury Triangulum Extended Region survey, we measured the star formation history near the position of 85 supernova remnants (SNRs) in M33. 
We constrained the progenitor masses for 60 of these SNRs, finding the remaining 25 remnants had no local star formation in the last 56 Myr consistent with core-collapse supernovae, making them potential Type Ia candidates.
We then infer a progenitor mass distribution from the age distribution, assuming single star evolution.
We find that the progenitor mass distribution is consistent with being drawn from a power-law with an index of \kstest.
Additionally, we infer a minimum progenitor mass of \bayesmass \M from this sample, consistent with several previous studies, providing further evidence that stars with ages older than the lifetimes of single 8 \M stars are producing supernovae.

\end{abstract}


\section{INTRODUCTION} \label{sect1}

Stellar evolution theory predicts that stars with a mass above $\sim$8 \M will end their lives as core-collapse supernovae (CCSNe; \citealt{Woosley02}).
The precise value of this lower limit has been the subject of multiple studies, some of which have found that red supergiants with masses \edit1{as low as} 7 \M \edit1{can be} progenitors for Type IIP SNe \citep{Smartt09,Fraser11,Jennings12}.

In addition to the uncertainty surrounding the lower mass limit of supernova (SN) progenitors, there has been growing evidence that not all stars with masses $>$8 \M experience a canonical \edit1{CCSNe} (e.g., \citealt{Smartt09,VD17}).
For example, stellar evolution theory suggests that there is an upper mass cutoff of $\sim$30 \M for Type II SNe \citep{Massey17}.
Observations, however, point to a new smaller upper limit of $17 - 19$ \M \citep{Smartt15}.
This discrepancy, along with the lack of high-mass progenitors in observations, was dubbed ``the red supergiant problem'' by \cite{Smartt09}, who first argued that the discrepancy is statistically significant.
More recently, \cite{DB20a} suggested that some of the discrepancy can be attributed to the steepness of the luminosity distribution of red supergiants as well as the small sample size.
They suggest the problem may not be as significant as previously believed, but also find a similar upper limit of $19^{+4}_{-2}$ $M_\odot$.
Above this mass limit, some stars may instead collapse straight into a black hole \citep{Pejcha15}.
These ``direct collapse'' systems would not produce a visible SN or leave behind a visible remnant.
The first observed black hole formation candidate occurred in the star-forming galaxy NGC 6946 \citep{Gerke15,Adams17}.
\cite{Murphy18} and \cite{Koplitz21} used the local stellar populations of the vanishing star to determine that the progenitor was likely $\sim$10.6 Myr old, which a single star progenitor points to an initial mass of $\sim$17 $M_\odot$.

Further progress on understanding the fates of massive stars requires increasing the number of CCSNe progenitors with mass constraints and expanding the measured distribution of progenitor masses to wider ranges of galaxy properties.
The traditional method for determining the mass of SNe progenitors is by directly imaging the progenitor stars (e.g. \citealt{Smartt03,Smartt04,VD03,Li06,GY07,Kilpatrick21}).
This technique requires high resolution (better than $\sim$0.$''$1) images of the SN site both before and after the event, which involves a large amount of serendipity.
The difficult requirement of having spatially resolved photometry of the location before the explosion has resulted in only 34 SNe having their progenitor masses determined by this method, along with \edit1{40} upper limits constrained (\citealt{VD17,Kilpatrick18,VD18,ONeill19,Kilpatrick21, Tinyanont2022,Vazquez2022}). % \citep{VD17,Kilpatrick18,VD18,ONeill19,Kilpatrick21,Vazquez2022,Tinyanont2022}
While the number of cataloged SNe has increased in recent years (e.g. \citealt{Guillochon17,Holoien19}), few of these SNe have had their progenitor constrained due to insufficient precursor imaging.

An alternative method, which does not require preexplosion images, uses an age-dating technique of the stellar populations surrounding an SN event \citep{Gogarten09,Murphy11}.
This technique leverages the stellar populations surrounding an SN to measure the local \edit1{star formation history (SFH)} by finding the model age distribution that best fits the color-magnitude diagram (CMD) of the resolved local stars.
By assuming the progenitor star belongs to the median population near the event, we are able to place statistical constraints on the age of the SN progenitor.
We can then infer the most likely mass of the progenitor by assuming that it was the most massive star that survives to that age according to the models.

This age-dating technique was shown to be a reliable way to infer progenitor ages for distances out to $\sim$8 Mpc \citep{Murphy11}.
Assuming only stars with masses $\gtrsim$7 \M become CCSNe requires photometry that is sensitive to populations as old as 56 Myr \citep{Girardi02}.
Because the technique does not require precursor imaging, it can be applied to any location where an SN has occurred in the recent past, including any known \edit1{SN remnants (SNRs)}.
As a result, several previous works have shown that most young stars within 50 pc of an SN event are associated with the progenitor \citep{Bastian06,Badenes09,Gogarten09,Jennings12,W14}.
For example, this technique was used to constrain the masses of SNR progenitors in the local star-forming galaxies M31 \citep{Jennings12}, NGC 6946 \citep{Koplitz21}, as well as the Magellanic Clouds \citep{Badenes09,Auchettl19}.
This technique has also been used to constrain the mass of observed CCSNe \citep{W14,W18,DR21,Koplitz21}.
Progenitor masses in M83 have also been constrained, including one with a most likely mass of 59 \M whose errors exclude ages older than 8 Myr, the highest mass progenitor inferred from the technique to date \citep{W19}.

M33, or the Triangulum Galaxy, is an excellent target for our technique.
It is nearby, relatively face on ($i = 56\degree$; \citealt{Zaritsky89}), and is known to host over 200 SNRs \citep{Long10,LL14}.
\cite{J14}, hereafter J14, have already applied this technique to 33 SNRs in M33, finding that the distribution was well fit by power-law distributions with indices that were significantly steeper than a standard Salpeter initial mass function power-law index of $-2.35$ \citep{Salpeter55}.
However, their analysis in M33 was limited by the heterogeneous set of archival Hubble Space Telescope (HST) images available.
This heterogeneous coverage resulted in inconsistent filter coverage and photometric depths between observations containing SNRs.
Furthermore, they did not fit a separate field star component to their ages, which could have resulted in age biases.
Here, we follow up on their work using the deep, uniform coverage provided by the \edit1{Panchromatic Hubble Andromeda Treasury Triangulum Extended Region (PHATTER)} survey \citep{phatter21} as well as updated fitting techniques.

The analysis we present here takes advantage of the work by \cite{DR18}, hereafter DR18, who developed a Bayesian hierarchical analysis capable of constraining the progenitor mass distribution index with an improved method for accounting for background effects as well as the minimum and maximum mass at which a star is able to undergo a CCSNe event from a set of SFHs.
They reanalyzed the SFHs from J14 as well as those from \cite{Lewis15} which correspond to likely SNRs from \cite{LeeLee14}, finding a progenitor mass index closer to, but not consistent with, a Salpeter index ($-2.96^{+0.45}_{-0.25}$).
This combined M31 and M33 distribution pointed to a minimum mass of $\sim$7.3 \M and a maximum mass of $>$59 $M_\odot$.
However, they found the SFHs from M31 led to a Salpeter progenitor mass distribution index ($-2.35^{+0.36}_{-0.48}$) with a minimum mass of 6.5 \M and a maximum mass of $>$46 $M_\odot$.

In this paper, we take an updated look at the ages of SNR progenitors in M33 using resolved stellar photometry from the PHATTER survey.
Our larger sample and more homogeneous photometry catalog allow us to compare different fitting methods and quantify the impact these changes have on the age and mass results.
Additionally, we compare our custom SNR-centered SFHs to those measured by \citet{Lazz22} in grids, allowing us to determine whether grid SFHs are sufficient for inferring a progenitor age and mass.
The rest of the paper is outlined as follows:
Section \ref{sect_data} details our SNR source catalog, as well as how our SFHs were measured.
\edit1{Section \ref{sect_res} presents our progenitor age and mass estimates.}
In Section \ref{sect_disc}, we discuss our constraint on the lower mass limit for CCSNe as well as the results of Kolmogorov $-$ Smirnov (KS) tests on our observed distribution, then compare our results to similar studies in the literature.
Finally, Section \ref{sect_sum} provides a short summary of our results. 
Throughout this paper, we assume a distance to M33 of 859 kpc \citep{deGrijs17}.

\section{DATA and ANALYSIS} \label{sect_data}

Our technique has two main data requirements. 
First, we need to know the locations of past SN activity.  
Second, we require resolved stellar photometry of the current populations within 50 pc of the SNe, as stars tend to remain spatially correlated within about 100 pc of their siblings for about 100 Myr, even if the cluster is not gravitationally bound \citep{Bastian06}.
Using these, we can measure the star formation rate as a function of lookback time, known as the SFH, at each SNR location.
The SFH provides the age distribution of the stars near each SN.
We then apply this age distribution to constrain the age and mass of the progenitor star.
We detail each of these steps below.

\subsection{SNR Locations} \label{sect_snrs}

For the locations of past SN activity in M33, we take the locations of SNRs from the catalogs of \cite{Long10} and \cite{LL14}, hereafter L10 and LL14, respectively.
L10 identified candidates based on their X-ray spectrum as well as having [\ion{S}{2}]:H$\alpha$ ratios $\geq$0.4.
The candidates in LL14 were identified based on their lack of blue stars, remnant morphology, and [\ion{S}{2}]:H$\alpha$ ratios $\geq$0.4.
Of the 137 SNR candidates in L10, 120 are included in LL14's catalog of 199 candidates.
The remaining 17 locations were classified as likely superbubbles or \ion{H}{2} regions, leading LL14 to exclude them from their final catalog.
Of these 17 potential SNRs, 4 (L10-043, L10-050, L10-079, L10-098) are within the PHATTER survey footprint.
We include these 4 locations in our catalog since they may be SNRs located within larger star forming complexes.
Of the 199 candidates from LL14, 81 reside in the PHATTER footprint, leading to our catalog of 85 SNR candidates.

In addition to the SNR locations, we produced \edit1{2} control \edit1{catalogs} of locations not associated with SNRs.
The first sample is 85 locations randomly distributed within the PHATTER footprint.
The second sample is 2500 random draws of the grid SFHs from \citet{Lazz22}, which do not contain an SNR.
Differences between the random and SNR samples provide additional evidence that the stellar populations near the SNRs are likely related to the progenitors, and not chance spatial coincidences of young stars (see Section \ref{sect_idx} for details).

\subsection{Photometry} \label{sect_sfh}

Once we had determined the historical SN locations, the second requirement was resolved stellar photometry at those locations.
This photometry was obtained from the PHATTER survey \citep{phatter21}.
The survey measured resolved stellar photometry for 22 million stars within M33 in optical (Advanced Camera for Surveys $F475W$ and $F814W$), near-ultraviolet (Wide Field Camera 3 (WFC3) $F275W$ and $F336W$), and near-infrared (WFC3 $F110W$ and $F160W$) bands.
Our photometry is derived from the optical images ($F475W$ and $F814W$) of the PHATTER survey, rather than measuring photometry in all 6 bands simultaneously.
We took samples from this photometry catalog for each SNR location and each control location, with the samples consisting of all of the stars within 50 pc ($12''$) from the SNR or random position.
We also collected samples of the widespread young populations surrounding each SNR from 50 to 1000 pc ($12''$ to $4'$).
These ``background'' samples allow us to identify young populations unique to the region containing the SNR.

To fit stellar evolution models to the photometric data, we require artificial star tests (ASTs) to correctly model the photometric completeness and uncertainty as a function of color and magnitude.
We used the ASTs from these data that were created by \cite{Lazz22}, who used them to measure grid SFHs in M33, as discussed in Section \ref{lazz_intro}.
These tests are obtained by adding stars of known flux to an image and blindly rerunning the photometry routine to measure the photometric bias, uncertainty, and completeness as a function of color and magnitude when fitting models to the data.
This is done at least 50,000 times within a region of interest.
\cite{W17} and \cite{Koplitz21} found that one set of artificial stars could be used for all locations of similar stellar density, rather than creating a set for each location.
This greatly reduces the computation time required.
\cite{Lazz22} used this technique to optimize the number of ASTs that needed to be created.
Since we are using the same photometry catalog as \cite{Lazz22}, we are able to use the same ASTs when analyzing the SNRs in our catalog.
These tests and the optical photometry catalog are described in further detail in \cite{Lazz22}.

\subsection{CMD Fitting}

Once we had the photometry and ASTs necessary to study each SNR location, we used the CMD fitting program MATCH \citep{dolphot02,dolphot12,dolphot13} to measure SFHs near the SNRs in our catalog.
MATCH has been used to constrain the age of SN progenitors (e.g., \citealt{Jennings12}; J14; \citealt{W18,W19,Koplitz21}) and SFH for nearby galaxies (e.g. \citealt{W09,Weisz2014,Skillman17}).
MATCH fits the observed CMD using the PARSEC stellar evolution models \citep{Bressan12}.
For each model age and metallicity, it creates a model CMD by assuming a Kroupa initial mass function \citep{Kroupa01}.
It then finds the highest likelihood linear combination of those models that provides the best fit to the observed CMD using a maximum likelihood estimator and taking into account the bias, uncertainty, and completeness of the photometry as determined by ASTs.
This combination of models yields the distribution of ages and metallicities for the stars in the observed CMD, which we refer to as the SFH of the region.

Below, we provide a brief description of our technique for running MATCH.
A more detailed account of the process can be found in \cite{Koplitz21}, which is identical to how we ran MATCH here.
In short, for each SNR location, we fit the CMD of the resolved stellar photometry with a grid of model CMDs generated from the PARSEC stellar evolution models \citep{Bressan12}.
Our model grid had time bins of size 0.05 dex from log$_{10}(t/\mathrm{yr}) = 6.6 - 8.0$ while bins of size 0.1 dex were used from log$_{10}(t/\mathrm{yr}) = 8.0 - 10.2$.
Since M33 is known to have a subsolar metallicity (e.g., \citealt{Barker11}), we limited the metallicities MATCH applies to the model grid to be $-0.5 \leq$ [Fe/H] $\leq 0.1$ using the $zinc$ flag.
Multiple massive stars can reside in the same location on CMDs even though they have different metallicities.
As a result, using the $zinc$ flag forces MATCH to use models for the young stars that are within the known metallicity range of M33.

As in \cite{Koplitz21}, our model also includes a ``background'' or ``contamination'' CMD of the stars in an annulus between 50 and 1000 pc ($12''$ $-$ $4'$) from the SNR.
The contamination CMD is scaled to the size of our regions before fitting.
This allows us to identify young populations that are sparse in the surrounding field and more heavily weight the populations concentrated within the regions being fit.

Furthermore, the fitting routine accounts for the effects of dust on the photometry.
We allowed MATCH to find the combination of reddening parameters along with the combination of ages and metallicities, which provided the best fit to the observed CMD. 
Since young populations are often found in dusty regions, MATCH applies three types of extinction to the model CMDs when fitting the stellar populations.
The first, A$_{V}$, is the total foreground extinction over the full region.
The second, dA$_{V}$, is the extinction spread due to the stars along the line of sight.
The third, dA$_{VY}$, is additional differential extinction added to populations younger than 100 Myr old.
The default dA$_{VY}$ value of 0.5 was used.
To determine A$_{V}$ and dA$_{V}$ for an SNR, we fit a range of possible values at the location.
We allowed A$_{V}$ to be between 0.05 and 1.00 in steps of 0.05 while dA$_{V}$ could be between 0.0 and 2.0 in steps of 0.2.

On average, our locations returned an A$_{V}$ value of 0.30, higher than the \cite{Schlafly11} value of 0.114.
This higher A$_{V}$ is not surprising given that MATCH takes into account the Milky Way and M33 reddening, whereas \cite{Schlafly11} only account for the Milky Way.
The vast majority of dA$_{V}$ values in our sample were 0.0, meaning the default differential reddening for the youngest stars (dA$_{VY} = 0.5$) was sufficient to account for differential reddening in most cases.

\subsection{Uncertainty Estimation}

Random and systematic uncertainties are inherent to fitting stellar models to CMDs.
Most of the random uncertainties in our fits arise from photometric errors as well as the number of stars used to determine the most likely progenitor age.
The systematic uncertainties are from any deficiencies present in the stellar evolution models used during the SFH fits.
\cite{Lazz22} have shown that there is good agreement between model sets for fits to young ages, and that the random uncertainties dominate the error budget in these fits.
Thus, we use the random uncertainties determination to estimate the uncertainties in our SFHs.

To estimate our random uncertainties, we used the \texttt{hybridMC} tool within MATCH \citep{dolphot13}.
This task uses a hybrid Monte Carlo algorithm to accept or reject potential SFHs around the best fit SFH based on likelihood.
We report the narrowest 68\% of the distribution of accepted SFHs that decreases with look-back time in columns (4) and (5) of Table \ref{tab_prob_ex}.
A detailed description of how our uncertainties are estimated can be found in Section \ref{sect_findmass}.

\subsection{SFHs from Previous Work} \label{lazz_intro}

Recently, \cite{Lazz22} published recent SFH maps of the PHATTER region of M33.
They used the same PHATTER optical photometry to measure the SFH of M33's inner disk in a grid of 100 $\times$ 100 pc ($24''$ $\times$ $24''$) cells, which they have released to the community.
\cite{Lazz22} largely used the same MATCH fitting technique as we have, but there were a few differences.
In their analysis, time bins of size 0.1 dex were used for all bins (log$_{10}(t/$yr$) = 6.6 - 10.2$).
Since they were measuring the total amount of star formation in each location, and not attempting to isolate very localized populations, a contamination CMD was not included during their fits.

Being able to constrain progenitor masses using such a grid of spatially resolved age distributions would be very powerful, since it would avoid having to access the original photometry and ASTs and run custom fitting for each SNR location.
Thus, we also attempted to age date the SNR locations using this grid of published SFHs by assigning an SFH from \cite{Lazz22} that corresponded to the location of each SNR in our sample.
We then compare the ages and masses of custom fits to those taken from a less optimized, but more easily accessible, source.

\subsection{Constraining Progenitor Mass} \label{sect_findmass}

The next step in constraining the masses of SNR progenitors is to convert the recent SFH from MATCH into a probability distribution for the age of the progenitor.
This calculation is done by determining the fraction of the total stellar mass present in each age bin.
We take this fraction to be equal to the probability that the progenitor is associated with that age.
We also take the error on that fraction as the error on the probability.
We provide an example of such a probability distribution in Table \ref{tab_prob_ex}.

The age probability distribution presented in Table \ref{tab_prob_ex} is for the SNR LL14-060.
Similar tables for each SNR with SF in the last 56 Myr are combined into one and made available in the online supplemental material.

While the age probability distribution derived from the SFH is the most complete constraint on the progenitor age, we also provide a single progenitor mass estimate with uncertainties.
This age simplifies the mass inference, as well as comparisons with other measurements and mass distribution analysis.
To derive the most likely progenitor age, we use the SFHs and uncertainties produced by MATCH to calculate the median age of the stellar populations younger than 56 Myr surrounding each SNR.
We then take that age as the most likely progenitor age.
We determine the uncertainties on the median age as follows.
We recalculate the median age a million times by accounting for the uncertainties and resampling the SF rates in each time bin, then determine the narrowest 68th percentile of this distribution of ages that contain the best fit. 
We use a 56 Myr cutoff for our SNR-centered SFHs, rather than the 50 Myr used by other works, because of the results of our Bayesian analysis presented in Section \ref{sect_bayes}.
A 56 Myr (log($t/\mathrm{yr}$) = 7.75) cutoff is not possible for the grid SFHs since \cite{Lazz22} ran MATCH with time bins of 0.1 dex.
As a result, we must decide whether to use a 50 or 63 Myr cutoff (log($t/\mathrm{yr}$) = 7.7 or 7.8) \edit1{for the grid SFH samples}.
We adopt a 50 Myr cutoff as this limits the number of contamination populations being included in our analysis.
To infer the progenitor mass for each age bin, we assume that the SNR progenitor is the highest surviving mass on the PARSEC stellar isochrone \citep{Bressan12}.

\begin{figure}
     \centering
     \epsscale{1.15}
     \plotone{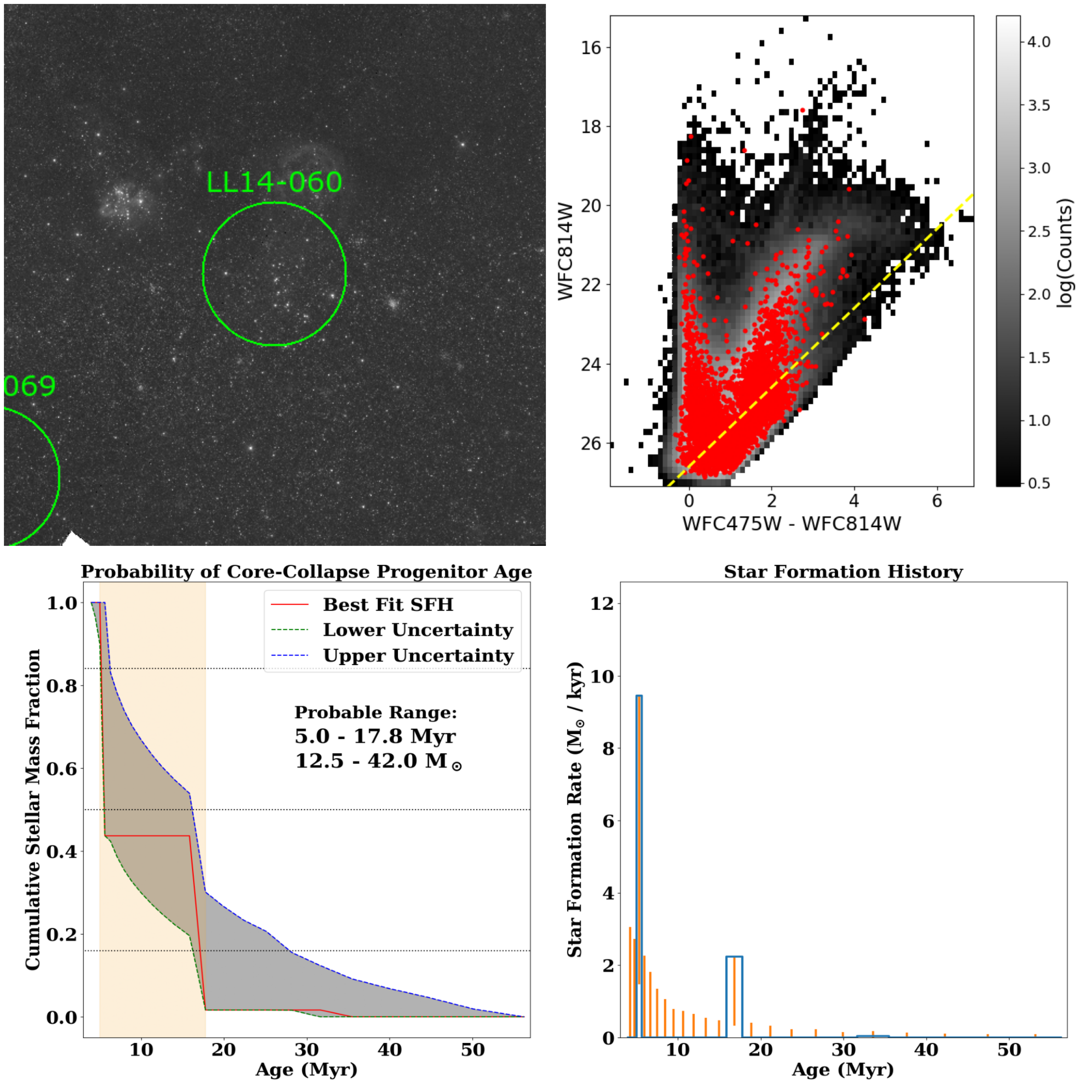}
     \caption{Data and analysis used to constrain the progenitor mass for the SNR LL14-060. \textit{Top left:} a $1.5' \times 1.5'$ $F475W$ image of the region of interest for the SNR with a green circle showing the 50 pc extraction region for the resolved stellar photometry sample. \textit{Top right:} observed CMD within our region of interest. Red points indicate the stars within 50 pc of the SNR while the the background field populations are plotted in gray scale and the magnitude limits of the data are shown as yellow dashed lines. \textit{Bottom left:} the best fit SFH and associated uncertainties produced by \texttt{hybridMC}. The cumulative fraction of stellar mass in each age bin is indicated by the red line. The gray shaded region depicts the 1$\sigma$ uncertainties on the SFH. The tan region shows the probable age range, which is taken to be median population of the 1$\sigma$ uncertainties. \textit{Bottom right:} the differential SFH, showing the SF rates and uncertainties that correspond to the cumulative fraction plot in \textit{bottom left}. The blue line indicates the SFH for the region, column (3) of Table \ref{tab_prob_ex}. The orange lines show the uncertainties in each time bin, column (3) minus column (4) to column (3) plus column (5) from Table \ref{tab_prob_ex}.
     \textit{We include figures of this format for all 60 SNRs with progenitor constraints as a figure set in the online Journal.}}
     \label{fig_example}
 \end{figure}

We present an example of our progenitor age fitting results for SNR LL14-160 in Figure \ref{fig_example}.
Similar summary plots are available for all of the SNR locations in the online supplemental material associated with the paper.

Past studies have shown that \edit1{progenitor masses estimated from the SFHs produced by MATCH are consistent with estimates from} other techniques (e.g., \citealt{Jennings12,W19,Koplitz21}).
DR18 found that their combined M31 and M33 distribution pointed to a minimum mass for CCSN progenitors of $\sim$7 $M_\odot$, which corresponds to an age of $\sim$50 Myr assuming single star evolution.
Populations older than this are more likely to be unrelated to the SNR since they have had more time to distance themselves from their parent cluster.
Older stars in \edit1{binaries} have been shown to be possible SN progenitors (e.g., \citealt{Xiao19}); however, our current inference from age to mass requires that we assume single star evolution.
Fortunately, this assumption should not impact our age constraints, which come from the surrounding population, but it could significantly impact our conversions between age and progenitor mass if the progenitor system was a mass-exchanging binary.

\section{Results} \label{sect_res}

We present and provide the progenitor mass results from our own custom SFHs.
We then compare to results that we obtain from previously published SFHs, as well as control samples and results from SNR studies of other nearby galaxies.
These comparisons suggest that custom SFHs \edit1{with a contamination CMD included in the fit to} account for the more widespread populations are required to isolate the ages of the stars most likely associated with each SNR.

\subsection{Comparing Grid SFHs to SNR-Centered SFHs} \label{sect_compare}

\begin{figure}
    \centering
    \plotone{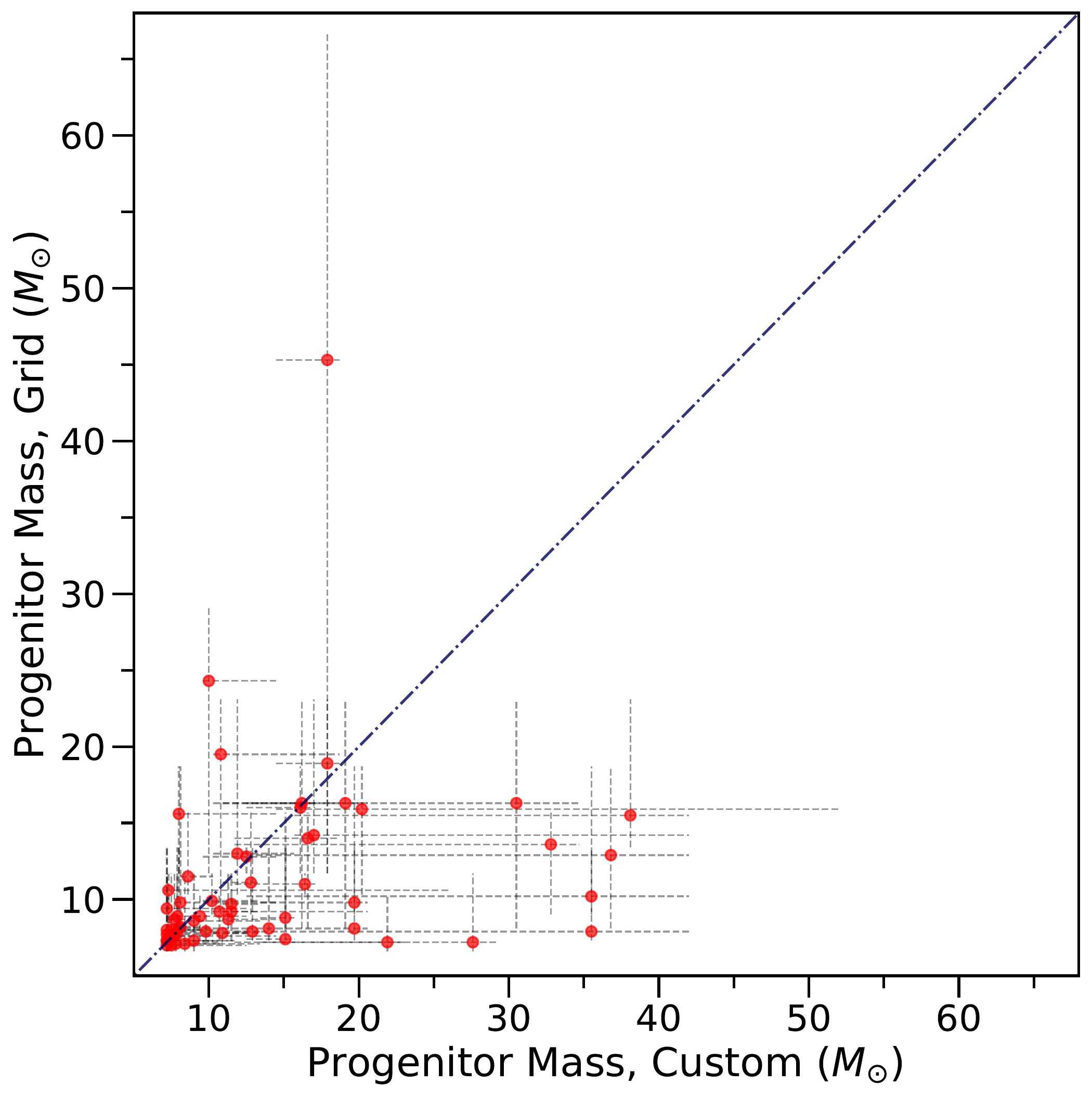}
    \caption{A comparison of the progenitor mass constraints for our sample derived from SNR-centered SFHs and grid SFHs \citep{Lazz22}. The red points indicate the progenitor mass estimates while the gray dashed lines show the associated uncertainties. The dark blue dashed-dotted line indicates the 1$-$to$-$1 line.}
    \label{fig_gc_comp}
\end{figure}

We present our progenitor mass constraints \edit1{for} the SNRs in our catalog in Table \ref{tab_results} \edit1{and} compare the resulting age distributions in Figure \ref{fig_gc_comp}, which reveals that the masses from \cite{Lazz22} are systematically lower than our custom measurements.
For 42 of the 85 locations in our catalog ($\sim$49\%) the best fit masses were not consistent with each other.
KS tests between these samples returned a $p-$value of 0.11, suggesting \edit1{we cannot rule out that they are} from the same parent distribution.

\begin{figure}
    \centering
    \plotone{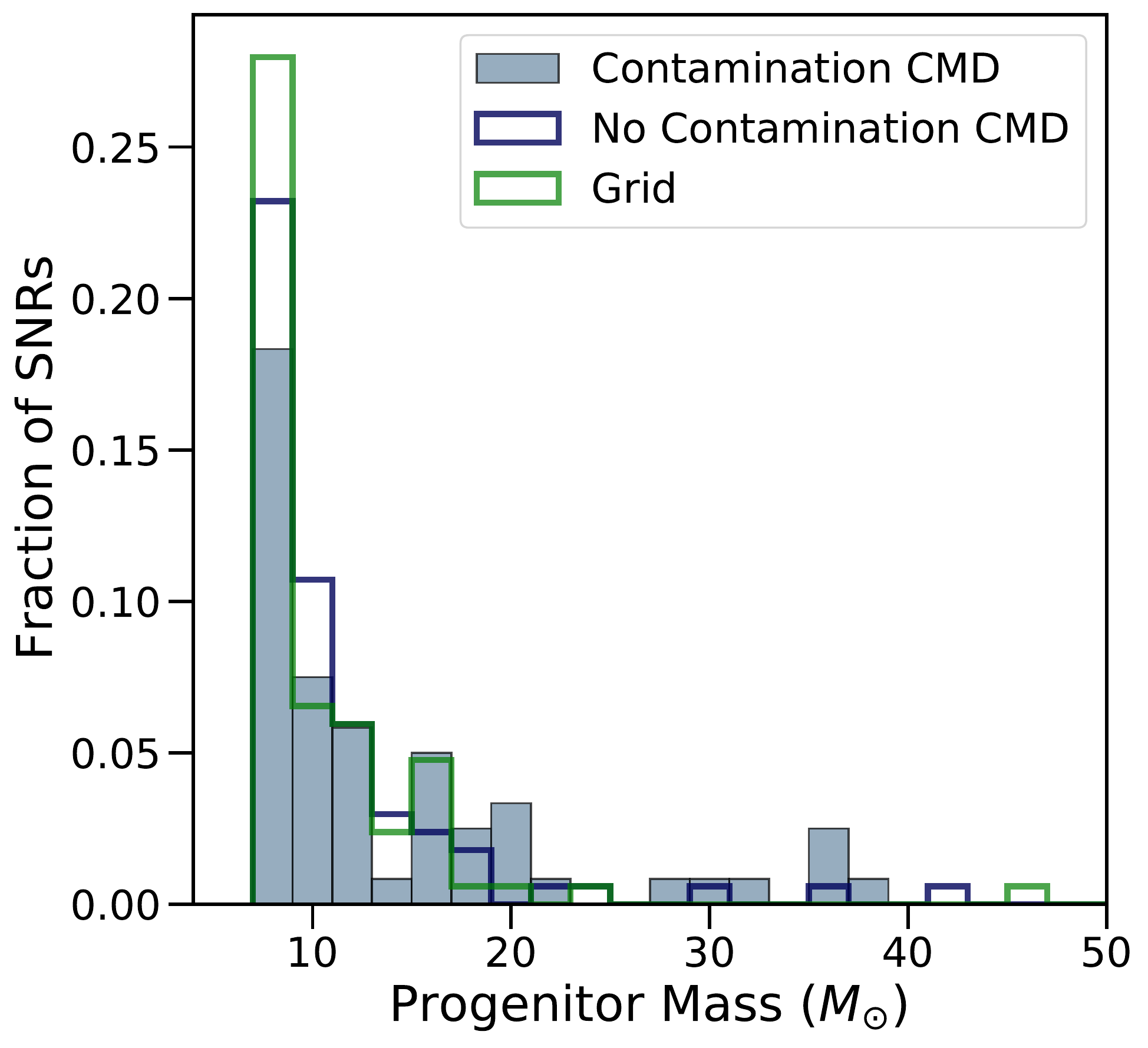}
    \caption{Histogram comparing the normalized progenitor mass distributions from \edit1{the} SNR-centered SFHs with and without a contamination CMD as well as the distribution from the grid SFHs \citep{Lazz22}.}% The grid SFHs and SNRs\edit1{-centered SFHs} without \edit1{a} contamination CMD resulted in only 1 Type Ia candidate (LL14-103). However, when a contamination CMD was used during the fits 25 locations were found to be Type Ia candidates. The lack of Ia candidates in the grid and no contamination CMD results is expected given that MATCH is not able to isolate populations unique to the extracted regions during these fits (see Section \ref{sect_sfh} for details).}
    \label{fig_hist_gc}
\end{figure}

\edit1{To isolate the cause of the observed difference, we reran our custom fits without including a contamination component, which returned a distribution similar to the one from the \cite{Lazz22} grid SFHs.
Performing KS tests between the SNR-centered distributions returned a $p-$value of 0.09 while 0.27 was returned when comparing the grid distribution to the SNR-centered without a contamination CMD sample.}
Figure \ref{fig_hist_gc} is a histogram comparing the distribution of progenitor masses that resulted from using the grid SFHs as well as the SNR-centered SFHs with and without a contamination CMD.
Each distribution is normalized such that they integrate to one.
The overall distribution from the grid SFHs is similar to that of our centered SFHs without a contamination CMD, which is expected given that both SFHs were fit without a contamination CMD and the sample populations overlap.

\edit1{Even though none of the distributions contain a progenitor that excludes masses $<$20 $M_\odot$, the grid SFHs produced 10 locations consistent with being more massive than 20 $M_\odot$ while the SNR-centered SFHs returned 15 with a contamination CMD and 9 without one.
A similar fraction of locations with masses between $7 - 15$ and $15 - 25$ \M were found in the grid and SNR-centered without a contamination component distributions ($86\%$, $13\%$ and $86\%$, $11\%$ respectively).
These show that the inclusion of the contamination CMD impacts the resulting distribution the most, though the high-precision custom location does play a large role.}
  
\subsection{Type Ia Candidates}  

Of the 85 locations in our catalog, \edit1{we classify 25 as Type Ia candidates.
\citet{Zapartas2017} showed that binaries with ages down to 200 Myr can produce delayed CCSNe; however, these systems have had enough time to move a significant distance away from their parent cluster, making the SF we measure older than $\sim$56 Myr likely contaminated by nearby populations that are not associated with the SN event.
Thus, any location without SF in the last $\sim$56 Myr we classify as Type Ia candidates since our technique cannot reliably determine the progenitor age beyond this.}

Including contamination CMDs in our SFH fits forces MATCH to only fit for SF above any \edit1{background} young stellar populations.
While this requirement can be helpful in isolating populations more likely to be associated with an SNR, it can also cause some SNRs to be classified as Type Ia candidates when they are actually Type II or Type Ibc in origin, because their associated young population may be too similar to that of the larger surroundings.
To check how many of our Type Ia candidates could actually be CCSNe, we can use our results from the \cite{Lazz22} SFHs, which measured the total star formation in each location.
The SNRs with mass estimates in \edit1{column (9) of Table \ref{tab_results}} but without a constraint in \edit1{column (7)} are less likely to be Type Ia in origin, as there are relatively high-mass populations nearby, just not above the background level.
Of the 25 Type Ia candidates from our SNR-centered SFHs, only LL14-103's grid SFH contained no SF within the last 50 Myr, making it our best Type Ia candidate.
The other 24 Type Ia candidates had young stellar populations present but not in sufficient quantities to be detected above the larger surroundings, making them weaker Type Ia candidates.
These results suggest a Type Ia fraction between 1 $-$ 29\%.
While this is not a tight constraint, it is consistent with the $\sim$15\% expected for late-type spirals \citep{Li11}.

\section{Discussion} \label{sect_disc}

Our progenitor age distributions probe the minimum mass at which CCSNe can occur, how SNe are spatially distributed in the disk of M33, and the power-law index of the progenitor mass distribution for the galaxy.

\subsection{Mass Limits for CCSNe} \label{sect_bayes}

\begin{figure}
    \centering
    \includegraphics[width=3.56in]{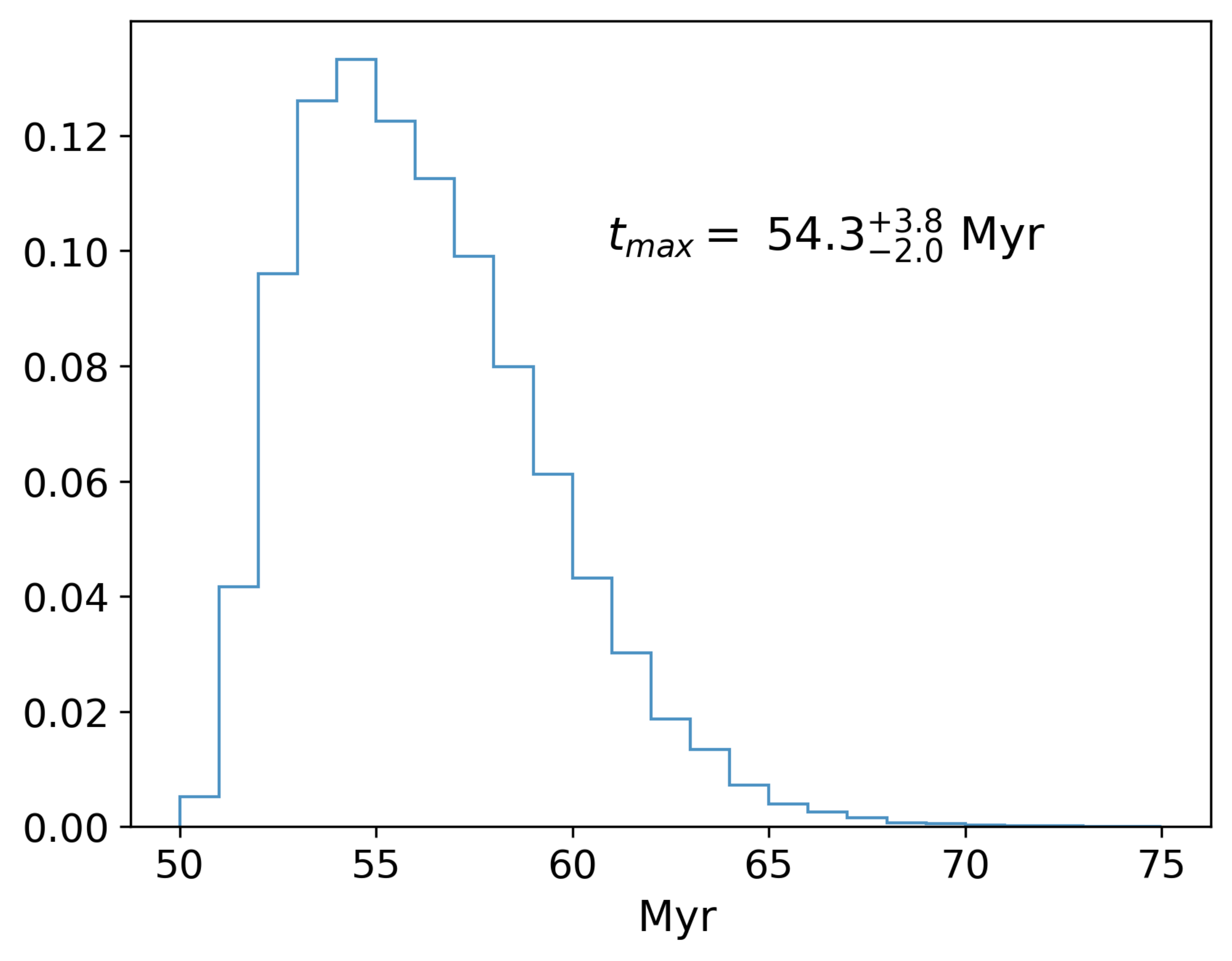}
    \includegraphics[width=3.5in]{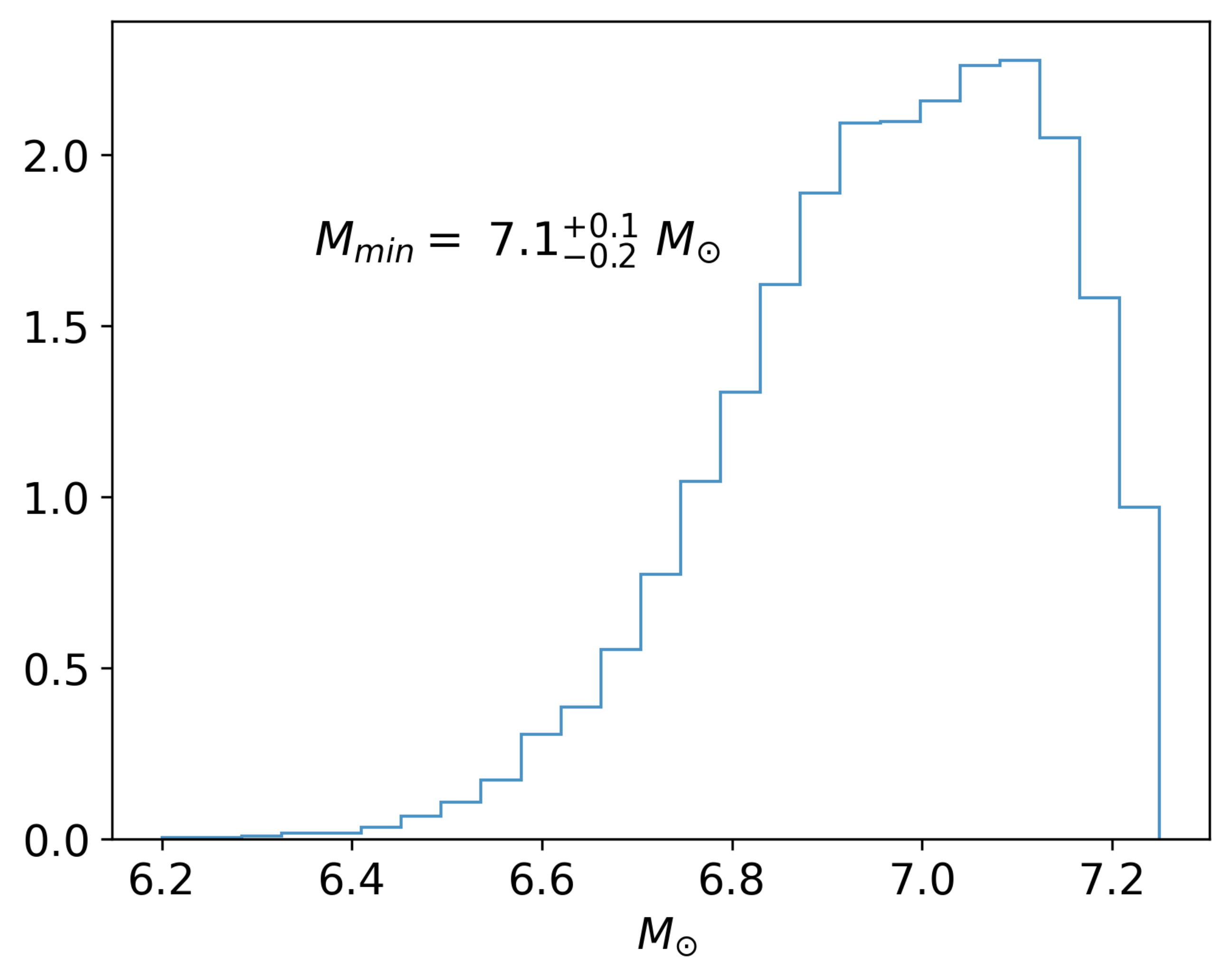}
    \caption{The results from performing a Bayesian hierarchical analysis on our full catalog of SNR progenitor ages. $Left$: the distribution of maximum ages at which stars undergo CCSNe, $t_\mathrm{max}$. $Right$: the distribution of minimum masses at which stars undergo CCSNe, $M_\mathrm{min}$. The $y$-axes in both histograms show the relative frequency of each age and mass. We report the fit assuming $t_\mathrm{min} = 15$ Myr, i.e., a minimum age for CCSNe of 15 Myr. This fit resulted in $t_\mathrm{max} =$ \bayesage Myr, corresponding to $M_\mathrm{min} =$ \bayesmass $M_{\odot}$. See Section \ref{sect_bayes} for details.}
    \label{fig_bayes}
\end{figure}

Using the Bayesian hierarchical analysis developed by DR18, \edit1{we use our SNR age sample} to provide a constraint on the maximum age at which stars undergo CCSNe, $t_\mathrm{max}$.
Our analysis was sensitive to the assumed minimum age for CCSNe, $t_\mathrm{min}$.
To account for this, we fit our distribution assuming $t_\mathrm{min}$ values of 6, 9, 10, 12, 15, and 18 Myr, with each returning similar results.
We report the $t_\mathrm{min} = 15$ Myr fit since this is the lowest $t_\mathrm{min}$ value that stabilized the returned progenitor mass distribution slope, finding \bayesage Myr as the best fit $t_\mathrm{max}$ which corresponds to a $M_\mathrm{min}$ of \bayesmass $M_\odot$.
Figure \ref{fig_bayes} shows the distribution of $t_\mathrm{max}$ (left) and $M_\mathrm{min}$ (right) returned by the Bayesian analysis for the fit with $t_\mathrm{min} = 15$ Myr.  

The analysis also attempts to constrain the upper mass \edit1{limit} for CCSNe and the progenitor mass distribution index when fitting a distribution.
Our high-mass progenitors, however, have large error bars that prevented the analysis from converging on a rigorous best fit value for the upper mass limit.
Since the upper mass is degenerate with the distribution index, it also did not return a reliable index.
We estimate the progenitor mass distribution index in Section \ref{sect_idx} using an alternate technique, since it may be of interest to the community.

\subsection{Spatial Distribution of Progenitor Masses}

\begin{figure}
    \centering
    \includegraphics[width=5.5in]{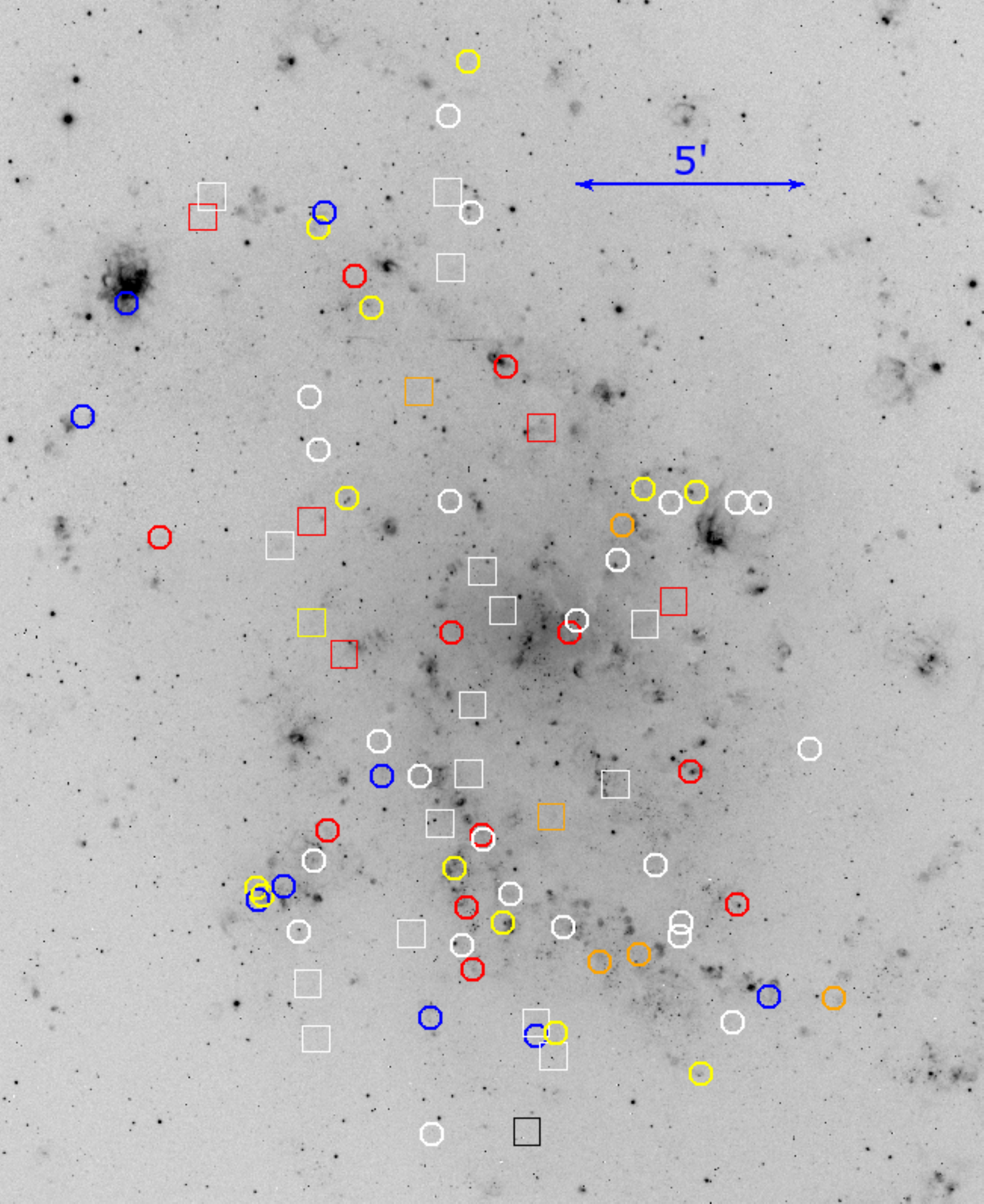}
    \includegraphics{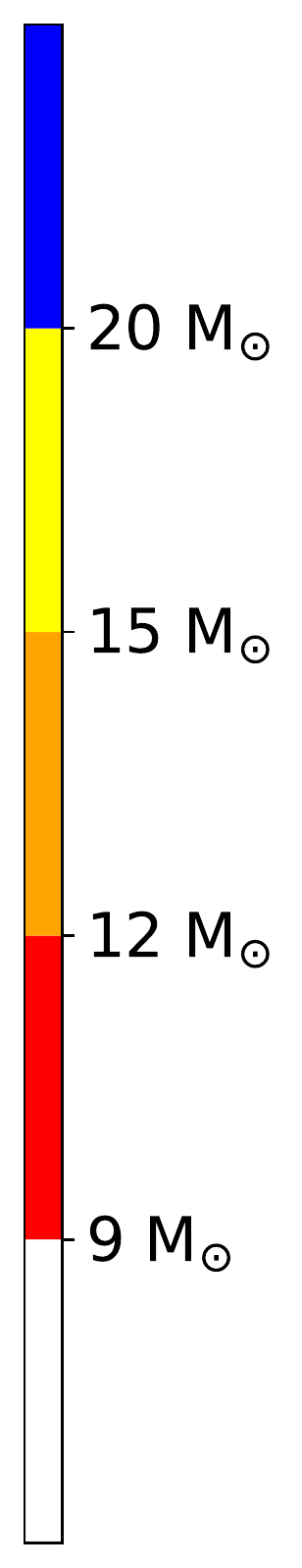}
    \caption{Locations of SNRs in M33 color coded by \edit1{their} progenitor masses overplotted on an H$\alpha$ image taken with the WIYN 0.9 m telescope \edit1{where the} mass and type of the progenitor are indicated by the color and symbol, respectively. SNRs with \edit1{an entry in column (7) of Table \ref{tab_results}} are shown as colored circles \edit1{with masses} $<$9 \M \edit1{in} white, masses of $9 - 12$ \M \edit1{in} red, masses of $12 - 15$ \M \edit1{in} orange, masses of $15 - 20$ \M \edit1{in} yellow, and masses $>$20 \M \edit1{in} blue. Type Ia candidates \edit1{(those without an entry in column (7) of Table \ref{tab_results})} are shown as colored squares, where the color indicates the \edit1{grid} progenitor mass \edit1{from column (9) of Table \ref{tab_results}}. The coloring is the same as the circles with the addition of black indicating the location of LL14-103, our best Type Ia candidate.}
    \label{fig_masscolor}
\end{figure}

To investigate the spatial distribution of SNRs in M33, we plot the locations of our catalog on an H$\alpha$ image taken with the WIYN 0.9m telescope (Figure \ref{fig_masscolor}).
The progenitor mass and most likely SNe type are indicated by the color and symbol, respectively.
Locations for which we have mass constraints in \edit1{column (7) of Table \ref{tab_results}}, i.e., we were able to measure SF within the last 56 Myr above the background level, are shown as circles.
Progenitors with masses $<$9 \M are white, masses of $9 - 12$ \M are red, masses of $12 - 15$\M are orange, masses of $15 - 20$ \M are yellow, and masses $>$20 \M are blue.
Our Type Ia SNe candidates are shown as squares, where the color indicates the best fit progenitor mass from the grid SFH that the SNR resides in from \cite{Lazz22}.
The colors show the mass that could have produced a CCSNe at the location, though these are less likely to be CCSNe than the colored circles due to the lower level of young SF.
The coloring depicts the same mass ranges as the circles, with the addition of black indicating \edit1{the location of} LL14-103, our best Type Ia candidate.

Our entire catalog mostly traces the H$\alpha$ emission and spiral arms of M33.
There are many squares (Type Ia candidates) inside star forming regions throughout the galaxy, indicating that young populations are present, just not enough to be detected in fits that include a contamination component.
In these cases, fitting without a contamination CMD (i.e., fitting the full population) often finds some massive stars in the region, whereas the fit including a contamination CMD finds no such populations.

\subsection{Progenitor Mass Distribution Power-Law Index} \label{sect_idx}

While the Bayesian hierarchical analysis of DR18 was not able to converge on a power-law index for the progenitor distribution due to the uncertainties at very young ages, it may still be of interest to determine the closest power-law representation of our most likely progenitor masses.
To determine this value, we use KS tests to determine the likelihood \edit1{the locations in} our catalog \edit1{with young populations ($<$56 Myr) are} drawn from various power-law distributions.
We compared the data to power-law indices between $-6.0$ and 0.0, in steps of 0.1, and report the most likely index.
\edit1{To estimate the uncertainties on the index, we employee a bootstrap analysis in which we sample the uncertainties on each mass 1000 times.
We then find the indices that return $p-$values $\geq$0.05 ($\sim$95\% confidence) and report the extremes as our limits.}

Performing this analysis on our full catalog of SNRs indicates the progenitor mass distribution is best matched by a power-law with an index of \kstest, which does contain the Salpeter index of $-2.35$ \citep{Salpeter55}.
The best fit index \edit1{has} a $p-$value of 0.23.
\edit1{Running} the progenitor mass distribution from the grid SFHs through this same analysis \edit1{found that the} sample was best matched by an index of $-3.8^{+1.8}_{-0.4}$, significantly steeper than our SNR-centered catalog \edit1{though still consistent}.
This is not surprising given that, as discussed in Section \ref{sect_compare}, L10-043 was the only progenitor found to be more massive than 25 $M_\odot$ \edit1{in this sample}.

\edit1{Our power-law indices were estimated using only the locations that contained SF at ages younger than $\sim$56 Myr.
To estimate the impact that removing locations without young SF has on our indices, we refit the progenitor mass distribution index of our SNR-centered sample while adding in the grid SFH progenitor mass for locations that did not contain young SF in our custom fit.
This combined sample was best fit by a $-3.2^{+1.3}_{-0.6}$ index, which is consistent with both the SNR-centered and grid SFH indices.
This indicates that removing locations without young SF does not have a large impact on the returned progenitor mass distribution index.}

\begin{figure}
    \centering
    \plotone{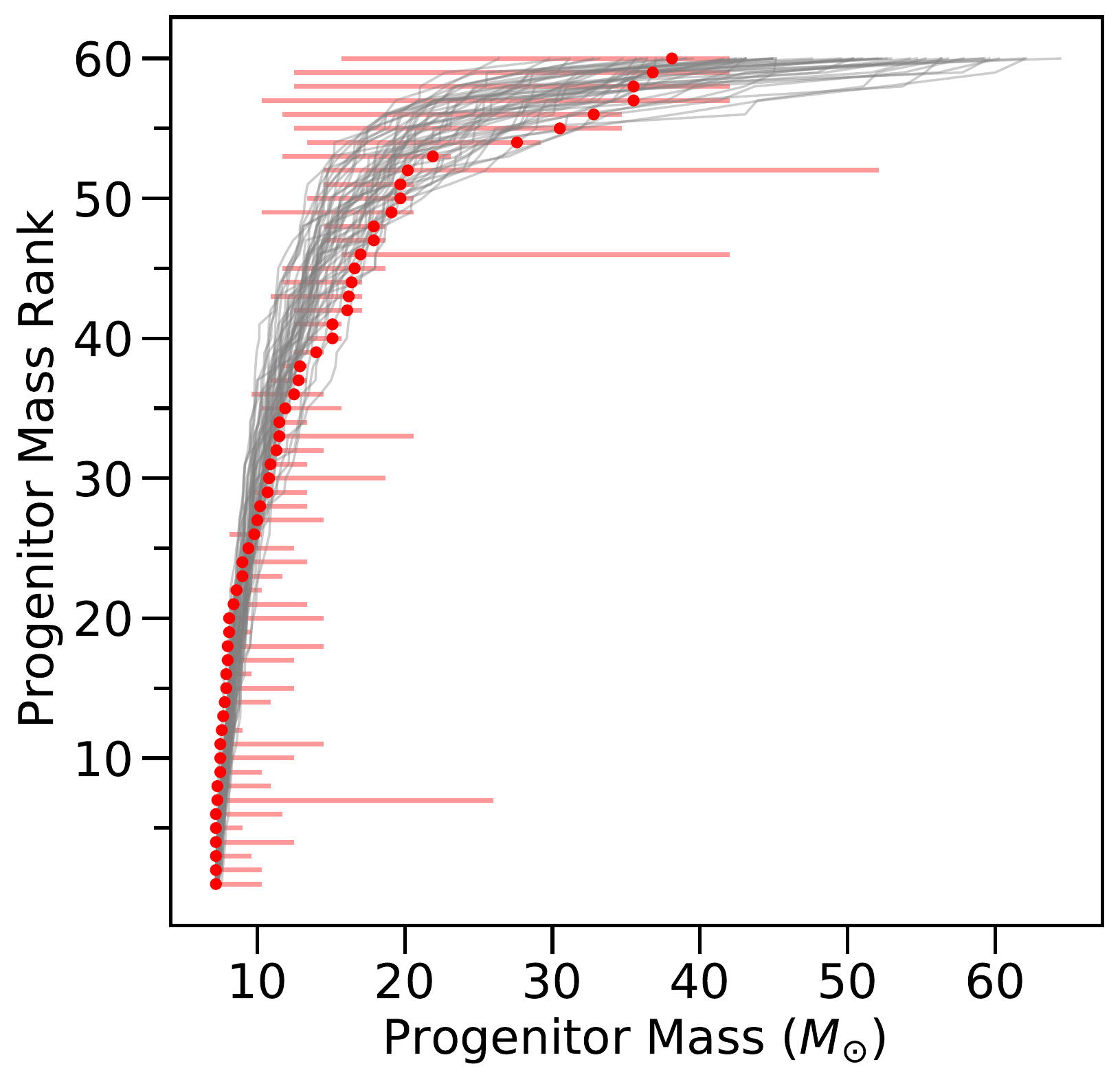}
    \caption{The rank distribution of progenitor masses for our full catalog of SNRs in M33. The progenitor masses are shown as red dots, with uncertainties depicted as red lines. Overplotted are 50 gray lines, which are draws from a power-law distribution with an index of \kstest, our best fit index.}
    \label{fig_rank}
\end{figure}

Figure \ref{fig_rank} shows our ranked progenitor mass distribution. 
The red points indicate the progenitor mass of each SNR with a constraint in \edit1{column (7) of Table \ref{tab_results}}, with uncertainties shown as red lines.
Overplotted as gray lines are 50 draws from a power-law distribution with an index of \kstest, our best fit index.

\subsection{Control Sample}

As mentioned in Section \ref{sect_snrs}, we also performed our analysis on control samples, random locations that did not contain SNRs.
\edit1{We compared our} SNR results to these control results to determine if the SNRs are indeed affecting our results.
We discuss both the control sample for the mass estimates based on custom SFH measurements and the control samples for mass estimates based on \cite{Lazz22} SFH measurements below.

Our first control sample, \edit1{containing randomly drawn locations in the PHATTER footprint}, returned fewer progenitors with masses $>$20 \M (5 in the control sample and 9 in our catalog).
There were also significantly more (\edit1{33, $\sim$39\% of the sample}) Type Ia candidates (i.e., locations with no significant recent SF above what is present in the contamination CMD) \edit1{than our SNR-centered distribution with a contamination CMD (25, $\sim$30\% of the sample)}.
Both of these suggest that the regions in this control sample contained, on average, older populations than those found near SNRs.
\edit1{The locations in this sample that contained SF at ages younger than 56 Myr were} best fit by a power-law index \edit1{of $-4.9^{+3.2}_{-0.2}$, which} is consistent with our contamination CMD and \edit1{the} grid distributions.
\edit1{While the uncertainties do overlap, this can likely be attributed to the amount of widespread SF within the PHATTER footprint of M33.
Comparing this random sample to our SNR-centered sample returned a $p-$value of 0.08, suggesting these sample are only marginally consistent with being drawn from the same parent distribution.}

Of the 2500 random draws in the grid control sample, $\sim$70\% were consistent with the grid power-law index, with the remaining $\sim$30\% resulting in steeper indices.
\edit1{We found the median index to be $-4.1 \pm 0.5$, which includes the grid SFH sample index of $-3.8$ but excludes the SNR-centered index of $-2.9$, though the uncertainties do overlap.
Of the 2500 draws, 1492 ($\sim$60\%) resulted in $p-$values $\leq$0.05 when compared to our grid sample.
No $p-$values $\geq$0.05 were found when compared to the SNR-centered sample.
Additionally, our grid sample only contained 1 location without recent SF (LL14-103) whereas only 3 of the grid control draws contained as many or fewer such locations,
meaning that $>$99\% of random draws had more locations without young stars present than we find in locations containing SNRs.}
These results show that the grid SFHs that contain SNRs do differ from those that lack an SNR, \edit1{with the similar power-law indices likely being explained, again, by the amount of widespread SF in M33}.

\subsection{Comparison to J14 and DR18} 

J14's catalog contained 33 SNRs in M33, of which 28 are in the PHATTER footprint.
Our SNR-centered progenitor mass estimates were consistent with those found by J14 for 16 of these 28 sources.
Of the 12 sources that were not consistent between our estimates and those in J14, we identify 8 Type Ia candidates.
\edit1{J14 found their full distribution of 33 SNRs was best fit by an index of $-3.8^{+0.5}_{-0.4}$.
Using our SNR-centered SFHs of the 28 overlapping locations, our analysis pointed to the distribution being well matched by a power-law index of $-3.1^{+1.2}_{-1.1}$, which is flatter than what J14 found.}
Their steep index \edit1{could be from} the few progenitors with masses $>$20 \M \edit1{in their sample, which} can likely be partially attributed to the low number of SNRs in their sample. 
\edit1{Poisson fluctuations for small numbers can randomly vary to zero quite easily.}
Additionally, J14 did not include a contamination CMD when fitting, which may have biased their estimates toward older, less massive populations and reduced their number of Type Ia candidates.
We have shown in Section \ref{sect_compare} that CMD-based age dating returns more low-mass progenitors when no background CMD is included in the fitting.
DR18 constrained the minimum mass for CCSNe to be $7.32^{+0.12}_{-0.14}$ $M_\odot$ using the J14 measurements, which is consistent with the minimum mass we identified, suggesting that this cutoff may be the most reliable parameter returned from the SNR sample.

\subsection{Comparison to Other Galaxies} \label{sect_othergals}

In addition to the SNRs in M33, J14 also constrained the progenitor mass distribution for 82 SNRs in M31 using photometry from the PHAT survey.
Their KS test analysis found that this distribution was best matched by a power-law index of $-4.4^{+0.4}_{-0.4}$, which is not consistent with our distribution.
Interestingly, one major difference between J14 and DR18 is that DR18 allow for a uniform background distribution when fitting for the progenitor mass distribution parameters.
This should have a similar \edit1{effect} to the use of a contamination CMD during the fitting process in that it accounts for the possibility that some of the measured SF may not be associated with the SN.
With the inclusion of the uniform background distribution, DR18 constrained the distribution index to be $-2.35^{+0.36}_{-0.48}$ and found the minimum mass for CCSNe to be $6.5^{+0.6}_{-0.2}$ \M using 62 SNRs in M31.
Both of \edit1{these measurements} are consistent with what we \edit1{have} found in M33.

\cite{Katsuda18} gathered the progenitor masses for 40 SNRs in the Milky Way and both Magellanic Clouds from the literature, which were estimated using chemical abundances.
They updated many of the measurements using Fe:Si ratios and found a progenitor mass distribution consistent with both a Salpeter index and our measured index for M33.

\cite{Auchettl19} examined 23 SNRs in the Small Magellanic Cloud, finding that 22 were likely core collapse in origin.
They report the likelihood that the mass of each progenitor is between $8-12.5$, $12.5-21.5$, and $>$21.5 $M_\odot$ assuming single and binary evolution.
Regardless of single or binary evolution, \edit1{70\%} of their progenitors had the highest likelihood in the most massive bin, whereas \edit1{9\%} of our SNRs were found to have progenitor masses in this range.
The large number of high-mass progenitors led to their distribution being well matched by a power-law index of $-1.84$, though the uncertainties are consistent with a Salpeter index.
A possible reason given by \cite{Auchettl19} for the top-heavy distribution is that the Small Magellanic Cloud has a lower metallicity than M33. 
It has been shown that lower metallicity gas is more likely to produce a top-heavy stellar distribution (e.g., \citealt{Bromm04,Marks12}).

\cite{W19} constrained the progenitor mass of 199 SNRs in M83 using our technique.
They found that the progenitor mass distribution was well matched by a power-law index of $-2.9^{+0.2}_{-0.7}$.
A KS test between their M83 distribution and ours resulted in a $p-$value of 0.04, suggesting their parent distributions may differ, \edit1{possibly due to the higher star formation intensity or lower metallicity of M33}.

\cite{Koplitz21} measured the progenitor mass of 169 SNRs, 8 historically observed SNe, and NGC6946-BH1, the first black hole formation candidate, in the galaxy NGC 6946 using our technique.
They found that gas emission impacted their broad $V$-band photometry, which biased some of their mass estimates.
As a result, they only included the 46 sources that were least likely to be biased when constraining the progenitor mass distribution index.
In this sample, they found \edit1{24\%} with masses $\geq$20 $M_\odot$, while we have \edit1{11\%} progenitors in our catalog with similar masses.
They found their distribution was best fit by an index of $-2.6^{+0.5}_{-0.6}$, which is consistent with our \edit1{measured} index.
KS tests between their preferred sample and our distribution resulted in a $p-$value of 0.2.

\begin{figure}
    \centering
    \plotone{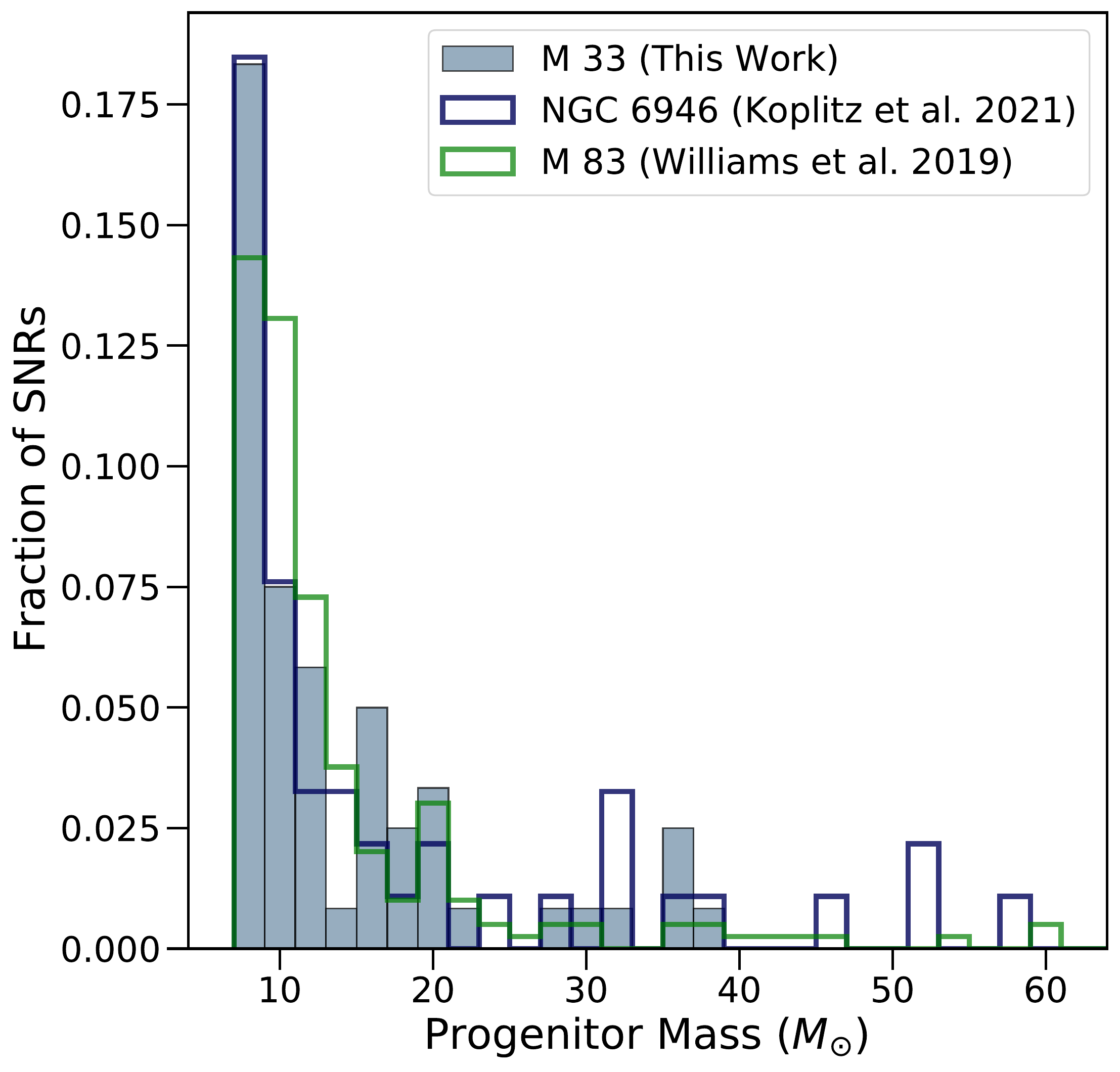}
    \caption{Histogram comparing the progenitor mass distribution of our catalog \edit1{(\kstest)} to the most reliable subsample in NGC 6946 \edit1{($-2.6^{+0.5}_{-0.6}$; \citealt{Koplitz21})} and the full catalog of SNRs in M83 \edit1{($-2.9^{+0.2}_{-0.7}$; \citealt{W19})}. Each bin is 2 \M wide and each distribution is normalized such that they integrate to one.}
    \label{fig_hist_catcomp}
\end{figure}

Figure \ref{fig_hist_catcomp} compares our distribution of progenitor masses to those in M83 \citep{W19} and the preferred sample in NGC 6946 \citep{Koplitz21}.
We normalize each individually so that they integrate to one.
Each is dominated by the low-mass progenitors and those less massive than 25 \M have similar overall shapes.
These led to power-law indices that are consistent with each other.
Our distribution is the only one that lacks any progenitors with mass $\geq$40 $M_\odot$.
However, this \edit1{could be the result of the small number of high-mass progenitors expected combined with} the smaller number of SNRs in \edit1{our} sample.

\section{Summary} \label{sect_sum}

We constrained the progenitor age and mass of 60 SNRs in the nearby galaxy M33, or the Triangulum Galaxy, using an age-dating technique of the stellar populations near the SNRs.
\edit1{The remaining 25 showed no local SF within the past 56 Myr, making them potential Type Ia candidates.
While it is possible that these candidates are binary systems producing delayed CCSNe with ages down to 200 Myr, our analysis is not able to reliably determine the progenitor age beyond $\sim$56 Myr.}

Using the Bayesian hierarchical analysis developed by DR18, we constrained the maximum age for CCSNe to be \bayesage Myr which, assuming single star evolution, corresponds to a minimum mass of \bayesmass $M_\odot$.
A KS test analysis determined that the progenitor mass distribution of our full catalog was best matched by a power-law distribution with an index of \kstest, which includes the Salpeter index \edit1{of} $-2.35$.
Our distribution is well populated by progenitors with masses \edit1{$9 - 40$} $M_\odot$. 

When using grid SFHs from \cite{Lazz22}, rather than SNR-centered regions with a contamination CMD included, the inferred progenitor mass was biased to lower values, with only 1 progenitor more massive than 25 $M_\odot$.
There were also fewer Type Ia candidates when using the grid SFHs, 1 from the grid SFHs and 25 from the SNR-centered SFHs.
Additionally, the progenitor mass distribution index that came from the grid SFHs was steeper than the index from the SNR-centered SFHs while KS tests between these samples returned a 0.03 $p-$value.
Without a contamination CMD, our custom SFHs returned a similar distribution to the grid, finding a $p-$value of 0.14 between the two samples.
The grid results differ from a random distribution of grid cells that do not contain an SNR, though not as strongly as our SNR-centered sample.
The stronger difference from the overall background suggests that the custom SFHs with a contamination CMD provide a more robust constraint on the age and mass of SNRs than SFHs measured in grids, where the background populations are not taken into account and the SNR may be anywhere in the grid cell.

Previously, J14 used archival HST images to constrain the age and mass of 33 SNRs in M33.
We present new age and mass estimates for 28 of these using the deep, uniform photometry from the PHATTER survey.
Performing KS test analysis on the SNRs with updated mass estimates pointed to the distribution being well matched by a power-law index of $-3.1^{+1.2}_{-1.1}$, which is consistent with the index J14 found for their full catalog \edit1{and our sample of 85 SNRs}.

Our normalized progenitor mass distribution is similar to that of M83 \citep{W19} and NGC 6946 \citep{Koplitz21}.
All the distribution are dominated by low-mass progenitors and have best fit power-law indices that are consistent with one another.
KS tests between our sample and the preferred sample in NGC 6946 resulted in a $p-$value of 0.20, suggesting that these are likely drawn from the same parent distribution. 
A $p-$value of 0.04 was returned when performing KS tests between our sample and the SNRs in M83.
\edit1{A $p-$value just below 0.05 and the matching power-law indices means we cannot rule out that the samples} are drawn from the same parent distribution.

Each of our distributions shows a sharp drop in the number of progenitors at $\sim$20 $M_\odot$.
Few progenitors are found more massive than this, which coincides with the upper limits found by \citet{Smartt15} and \citet{DB20a} for Type II SNe.
It is possible that the reason we do not see many high mass progenitors is because not all experience a canonical CCSNe and instead collapse directly into a black hole \citep{Pejcha15}.
Now that the JWST has launched, similar studies will be able to leverage red supergiants to constrain the age of SN progenitors with higher precision and may resolve the ``red supergiant problem''.

Support for this work was provided by NASA through grants GO-14610 and GO-15216 from the Space Telescope Science Institute, which is operated by AURA, Inc., under NASA contract NAS 5-26555.

\bibliographystyle{aasjournal}
\bibliography{References}

\begin{deluxetable}{ccccccccccccc} 
\label{tab_prob_ex} 
\centering 
\tablewidth{0pt} 
\tablecolumns{13} 
\tablecaption{Age Probability Distribution Results for LL14-060} 
\tablehead{
\colhead{T1} & \colhead{T2} & \colhead{SFR(Best)} & \colhead{-err} & \colhead{+err} & \colhead{PDF(Best)} & \colhead{-err} & \colhead{+err} & \colhead{CDF(Best)} & \colhead{CDF(Low)} & \colhead{CDF(High)} & \colhead{M1} & \colhead{M2}\\
(1) & (2) & (3) & (4) & (5) & (6) & (7) & (8) & (9) & (10) & (11) & (12) & (13)}
\startdata
  4.0 &   4.5 &  0.0000e+00 &  0.0000e+00 &  3.0515e-03 &     0.000 &  0.000 &  0.263 &     0.000 &    0.000 &     0.035 &  52.1 &  66.6 \\
  4.5 &   5.0 &  0.0000e+00 &  0.0000e+00 &  2.7223e-03 &     0.000 &  0.000 &  0.264 &     0.000 &    0.000 &     0.101 &  42.0 &  52.1 \\
  5.0 &   5.6 &  9.4464e-03 &  7.9741e-03 &  1.5900e-05 &     0.563 &  0.471 &  0.316 &     0.563 &    0.000 &     0.563 &  34.7 &  42.0 \\
  5.6 &   6.3 &  0.0000e+00 &  0.0000e+00 &  2.2637e-03 &     0.000 &  0.000 &  0.273 &     0.563 &    0.169 &     0.576 &  29.2 &  34.7 \\
  6.3 &   7.1 &  0.0000e+00 &  0.0000e+00 &  1.8153e-03 &     0.000 &  0.000 &  0.252 &     0.563 &    0.218 &     0.613 &  26.0 &  29.2 \\
  7.1 &   7.9 &  0.0000e+00 &  0.0000e+00 &  1.3441e-03 &     0.000 &  0.000 &  0.219 &     0.563 &    0.260 &     0.645 &  23.1 &  26.0 \\
  7.9 &   8.9 &  0.0000e+00 &  0.0000e+00 &  1.0684e-03 &     0.000 &  0.000 &  0.200 &     0.563 &    0.298 &     0.674 &  20.6 &  23.1 \\
  8.9 &  10.0 &  0.0000e+00 &  0.0000e+00 &  7.8472e-04 &     0.000 &  0.000 &  0.171 &     0.563 &    0.332 &     0.701 &  18.7 &  20.6 \\
 10.0 &  11.2 &  0.0000e+00 &  0.0000e+00 &  7.4253e-04 &     0.000 &  0.000 &  0.179 &     0.563 &    0.365 &     0.727 &  17.1 &  18.7 \\
 11.2 &  12.6 &  0.0000e+00 &  0.0000e+00 &  6.5002e-04 &     0.000 &  0.000 &  0.177 &     0.563 &    0.398 &     0.753 &  15.7 &  17.1 \\
 12.6 &  14.1 &  0.0000e+00 &  0.0000e+00 &  5.4978e-04 &     0.000 &  0.000 &  0.169 &     0.563 &    0.430 &     0.779 &  14.5 &  15.7 \\
 14.1 &  15.8 &  0.0000e+00 &  0.0000e+00 &  4.8213e-04 &     0.000 &  0.000 &  0.167 &     0.563 &    0.461 &     0.804 &  13.4 &  14.5 \\
 15.8 &  17.8 &  2.2299e-03 &  1.9046e-03 &  0.0000e+00 &     0.420 &  0.363 &  0.381 &     0.983 &    0.699 &     0.983 &  12.5 &  13.4 \\
 17.8 &  20.0 &  0.0000e+00 &  0.0000e+00 &  4.0795e-04 &     0.000 &  0.000 &  0.176 &     0.983 &    0.734 &     0.983 &  11.7 &  12.5 \\
 20.0 &  22.4 &  0.0000e+00 &  0.0000e+00 &  3.2564e-04 &     0.000 &  0.000 &  0.161 &     0.983 &    0.767 &     0.983 &  10.9 &  11.7 \\
 22.4 &  25.1 &  0.0000e+00 &  0.0000e+00 &  2.3611e-04 &     0.000 &  0.000 &  0.135 &     0.983 &    0.794 &     0.983 &  10.3 &  10.9 \\
 25.1 &  28.2 &  0.0000e+00 &  0.0000e+00 &  2.2838e-04 &     0.000 &  0.000 &  0.144 &     0.983 &    0.845 &     0.984 &   9.6 &  10.3 \\
 28.2 &  31.6 &  0.0000e+00 &  0.0000e+00 &  1.5449e-04 &     0.000 &  0.000 &  0.114 &     0.983 &    0.876 &     1.000 &   9.0 &   9.6 \\
 31.6 &  35.5 &  4.3935e-05 &  4.3935e-05 &  1.3031e-04 &     0.017 &  0.017 &  0.128 &     1.000 &    0.908 &     1.000 &   8.6 &   9.0 \\
 35.5 &  39.8 &  0.0000e+00 &  0.0000e+00 &  1.2988e-04 &     0.000 &  0.000 &  0.119 &     1.000 &    0.931 &     1.000 &   8.1 &   8.6 \\
 39.8 &  44.7 &  0.0000e+00 &  0.0000e+00 &  1.1090e-04 &     0.000 &  0.000 &  0.115 &     1.000 &    0.953 &     1.000 &   7.7 &   8.1 \\
 44.7 &  50.1 &  0.0000e+00 &  0.0000e+00 &  9.9118e-05 &     0.000 &  0.000 &  0.115 &     1.000 &    0.981 &     1.000 &   7.3 &   7.7 \\
 50.1 &  56.2 &  0.0000e+00 &  0.0000e+00 &  9.1503e-05 &     0.000 &  0.000 &  0.119 &     1.000 &    1.000 &     1.000 &   7.0 &   7.3 \\
 \enddata
\tablenotetext{}{Columns (1) and (2) define the beginning and end of the age bin in log$_{10}(t/$yr$)$. Column (3) is the best fit SF rate in that age bin in \M yr$^{-1}$. Columns (4) and (5) are the lower and upper limits on that rate. Column (6) shows the fraction of stellar mass younger than 56 Myr in the age bin according to the best fit SFH equivalent to the probability of that progenitor age. Columns (7) and (8) are the negative and positive error bars on this probability. Column (9) provides the cumulative fraction of stellar mass over time for the best fit SFH. Columns (10) and (11) are the lower and upper uncertainties of the cumulative mass fraction for a set of a million realizations of the SFH with the uncertainties in columns (4) and (5). Columns (12) and (13) give the mass of stars with lifetimes within the age bin in $M_\odot$.}
\tablecomments{Table \ref{tab_prob_ex} is published in its entirety in the machine-readable format. A portion is shown here for guidance regarding its form and content.}
\end{deluxetable}

\begin{deluxetable}{ccccccccc} 
\label{tab_results} 
\centering 
\tablewidth{0pt} 
\tablecolumns{9} 
\tablecaption{M33 Supernova Remnants} 
\tablehead{
\colhead{ID$^{a,b}$} & \colhead{R.A. (J2000)$^{a,b}$} & \colhead{Decl. (J2000)$^{a,b}$} & \colhead{Stars} & \colhead{dA$_{V}$} & \colhead{A$_{V}$} & \colhead{Mass ($M_{\odot}$)} & \colhead{Region ID$^c$} & \colhead{Grid Mass ($M_{\odot}$)}\\
(1) & (2) & (3) & (4) & (5) & (6) & (7) & (8) & (9)} 
\startdata 
L10-043 & 01:33:35.39 & +30:42:32.4 & 5510 & 0.00 & 0.45 & $17.9^{+0.8}_{-3.4}$ & 1206 & $45.3^{+21.3}_{-33.6}$\\
L10-050 & 01:33:40.73 & +30:42:35.7 & 7591 & 0.00 & 0.30 & $16.6^{+2.1}_{-4.9}$ & 1208 & $14.0^{+1.7}_{-5.9}$\\
L10-079 & 01:33:58.15 & +30:48:37.2 & 5115 & 0.00 & 0.35 & $9.0^{+4.4}_{-0.4}$ & 1791 & $8.6^{+3.1}_{-0.5}$\\
L10-098 & 01:34:12.69 & +30:35:12.0 & 5334 & 0.00 & 0.40 & $10.8^{+7.9}_{-0.5}$ & 0534 & $19.5^{+3.6}_{-10.5}$\\
LL14-050 & 01:33:21.56 & +30:31:31.1 & 3471 & 0.40 & 0.40 & $14.0^{+0.5}_{-1.5}$ & 0178 & $8.1^{+5.3}_{-0.8}$\\
LL14-054 & 01:33:24.01 & +30:36:56.8 & 5599 & 0.00 & 0.25 & $7.3^{+18.7}_{-0.3}$ & 0686 & $10.6^{+1.1}_{-2.5}$\\
LL14-060 & 01:33:28.10 & +30:31:35.0 & 3965 & 0.00 & 0.40 & $35.5^{+6.5}_{-23.0}$ & 0182 & $10.2^{+3.2}_{-1.2}$\\
LL14-061 & 01:33:29.04 & +30:42:17.3 & 5522 & 0.00 & 0.40 & $7.2^{+1.8}_{-0.2}$ & 1163 & $7.0^{+6.4}_{-0.4}$\\
LL14-067 & 01:33:31.38 & +30:33:33.4 & 6592 & 0.00 & 0.15 & $9.8^{+0.5}_{-1.7}$ & 0365 & $7.9^{+1.1}_{-0.6}$\\
LL14-068 & 01:33:31.32 & +30:42:18.3 & 6023 & 0.00 & 0.30 & $7.2^{+4.5}_{-0.2}$ & 1164 & $7.3^{+0.8}_{-0.7}$\\
... & ... & ... & ... & ... & ... & ... & ... & ...\\
\enddata 
\tablenotetext{}{Column (1) is the SNR identifier from L10 or LL14. Columns (2) and (3) are the R.A. and Decl. (J2000) of the SNR. Column (4) indicates the number of stars that were used to measure the SFH. Columns (5) and (6) are the best fit dA$_{V}$ and A$_{V}$ values, respectively. Column (7) is the progenitor mass inferred from the median age calculated from the best fit SFH, and associated uncertainties in $M_\odot$. Column (8) indicates which region from \cite{Lazz22} the SNR is associated with. Column (9) is the progenitor mass and uncertainties that resulted from the grid SFH.}
\tablenotetext{a}{\cite{Long10}} 
\tablenotetext{b}{\cite{LL14}} 
\tablenotetext{c}{\cite{Lazz22}} 
\tablecomments{(This table is available in its entirety in machine-readable form.)}
\end{deluxetable} 

\end{document}